\documentclass[twocolumn]{aastex631}

\usepackage{style}

\begin{document}

\title{The NANOGrav 15 yr Data Set: Running of the Spectral Index}
\author[0000-0001-5134-3925]{Gabriella Agazie}
\affiliation{Center for Gravitation, Cosmology and Astrophysics, Department of Physics, University of Wisconsin-Milwaukee,\\ P.O. Box 413, Milwaukee, WI 53201, USA}
\author[0000-0002-8935-9882]{Akash Anumarlapudi}
\affiliation{Center for Gravitation, Cosmology and Astrophysics, Department of Physics, University of Wisconsin-Milwaukee,\\ P.O. Box 413, Milwaukee, WI 53201, USA}
\author[0000-0003-0638-3340]{Anne M. Archibald}
\affiliation{Newcastle University, NE1 7RU, UK}
\author{Zaven Arzoumanian}
\affiliation{X-Ray Astrophysics Laboratory, NASA Goddard Space Flight Center, Code 662, Greenbelt, MD 20771, USA}
\author[0000-0002-4972-1525]{Jeremy G.\ Baier}
\affiliation{Department of Physics, Oregon State University, Corvallis, OR 97331, USA}
\author[0000-0003-2745-753X]{Paul T. Baker}
\affiliation{Department of Physics and Astronomy, Widener University, One University Place, Chester, PA 19013, USA}
\author[0000-0003-0909-5563]{Bence B\'{e}csy}
\affiliation{Department of Physics, Oregon State University, Corvallis, OR 97331, USA}
\author[0000-0002-2183-1087]{Laura Blecha}
\affiliation{Physics Department, University of Florida, Gainesville, FL 32611, USA}
\author[0000-0001-6341-7178]{Adam Brazier}
\affiliation{Cornell Center for Astrophysics and Planetary Science and Department of Astronomy, Cornell University, Ithaca, NY 14853, USA}
\affiliation{Cornell Center for Advanced Computing, Cornell University, Ithaca, NY 14853, USA}
\author[0000-0003-3053-6538]{Paul R. Brook}
\affiliation{Institute for Gravitational Wave Astronomy and School of Physics and Astronomy, University of Birmingham, Edgbaston, Birmingham B15 2TT, UK}
% ...
\shorttitle{Running of the Spectral Index}
\shortauthors{The NANOGrav Collaboration}

\correspondingauthor{Rafael R.\ Lino dos Santos}
\email{rafaellinodossantos@nanograv.org}

\correspondingauthor{Kai Schmitz}
\email{kai.schmitz@nanograv.org}

\begin{abstract}
The NANOGrav 15-year data provides compelling evidence for a stochastic gravitational-wave (GW) background at nanohertz frequencies. The simplest model-independent approach to characterizing the frequency spectrum of this signal consists in a simple power-law fit involving two parameters: an amplitude $A$ and a spectral index $\gamma$. In this paper, we consider the next logical step beyond this minimal spectral model, allowing for a \textit{running} (i.e., logarithmic frequency dependence) of the spectral index, $\gamma_{\rm run}(f) = \gamma + \beta \ln\left(f/f_{\rm ref}\right)$. We fit this running-power-law (RPL) model to the NANOGrav 15-year data and perform a Bayesian model comparison with the minimal constant-power-law (CPL) model, which results in a 95\,\% credible interval for the parameter $\beta$ consistent with no running, $\beta \in \left[-0.80,2.96\right]$, and an inconclusive Bayes factor, $\mathcal{B}\left(\textrm{RPL~vs.~CPL}\right) = 0.69 \pm 0.01$. We thus conclude that, at present, the minimal CPL model still suffices to adequately describe the NANOGrav signal; however, future data sets may well lead to a measurement of nonzero $\beta$. Finally, we interpret the RPL model as a description of primordial GWs generated during cosmic inflation, which allows us to combine our results with upper limits from big-bang nucleosynthesis, the cosmic microwave background, and LIGO--Virgo--KAGRA.
\end{abstract}




\section{Introduction} 
\label{sec:introduction}

Pulsar timing arrays (PTAs) are gravitational-wave (GW) detectors of galactic dimensions that aim to measure the imprint of a nanohertz GW background (GWB) in the timing data of millisecond pulsars~\citep{Taylor:2021yjx}. Recently, the field of PTA searches for GWs reached an important milestone when the CPTA~\citep{Xu:2023wog}, EPTA and InPTA~\citep{EPTA:2023fyk}, NANOGrav~\citep{NANOGrav:2023gor}, and PPTA~\citep{Reardon:2023gzh} Collaborations announced the first observational evidence for the Hellings--Downs (HD) curve~\citep{Hellings:1983fr}, i.e., the characteristic cross-correlation pattern that general relativity predicts a GWB to induce in the timing-residual cross-power spectrum for pairs of pulsars in the sky. The NANOGrav 15-year (NG15) data~\citep{NANOGrav:2023hde} in particular provides compelling evidence for the presence of an HD-correlated common-spectrum process (i.e., a GWB) at nanohertz frequencies, at a level of statistical significance of $3.5 \cdots 4.0\,\sigma$~\citep{NANOGrav:2023gor}.

Assuming the signal in the NG15 data to correspond to a genuine GWB, one is faced with the question of its origin. The most common expectation is that the signal is caused by a cosmic population of inspiraling supermassive black-hole binaries (SMBHBs) at the centers of galaxies~\citep{NANOGrav:2023hfp}. Alternatively, it may represent a GW echo of the big bang, i.e., a primordial GWB signal produced by new particle physics in the very early Universe~\citep{Caprini:2018mtu,NANOGrav:2023hvm,EPTA:2023xxk}. In order to resolve this dichotomy and pin down the origin of the signal, more work is needed. In the coming years, searches for continuous-wave signals~\citep{NANOGrav:2023pdq} and GWB anisotropies~\citep{NANOGrav:2023tcn} promise to shed more light on the origin of the signal. However, for the time being, model selection mostly relies on the spectral characterization of the signal~\citep{Lamb:2023jls,Mitridate:2023oar,Gersbach:2024hcc}\,---\,which is what we will be concerned with in this paper. 

Explicit astrophysical and cosmological models often yield specific predictions for the spectral shape of the GWB. For instance, the simplest SMBHB models, in which binary evolution is purely driven by GW emission, predict a power-law shape with a characteristic spectral index, $\gamma \simeq 13/3$ (see below for the definition of $\gamma$), in the limit of a large number of sources~\citep{Phinney:2001di}. The GW signal from a first-order phase transition in the early Universe, on the other hand, is expected to have the shape of a broken power law or even a doubly broken power law~\citep{Caprini:2024hue}. These predictions are representative of the top-down approach to the spectral characterization of the signal, i.e., the idea to first start from a concrete physical model (possibly involving physics at very high energies) and then work out the observational consequences in the PTA frequency band. In parallel, it is imperative to develop model-independent spectral templates that enable a bottom-up approach to the spectral characterization of the signal, i.e., an approach that starts with an agnostic description of features in the data and then asks which GWB models might be able to account for these features. At present, two model-independent templates are commonly used in the PTA literature: (i) a simple power law, parametrized in terms of an amplitude $A$ (at some reference frequency $f_{\rm ref}$) and a spectral index $\gamma$; and (ii) a free spectrum, which treats the GWB amplitude in each frequency bin as a free parameter. The purpose of the present paper is to extend this list of templates by a third one. 

The power-law template clearly represents the simplest model-independent ansatz for the spectral shape of the GWB. The spectral index $\gamma$ in this model is assumed to be constant, which is why we will refer to this model also as the constant-power-law (CPL) model in the following. If plotted on doubly logarithmic axes, the function graph of a power law is nothing but a straight line. Therefore, if we seek to construct next-to-minimal GWB templates, the next logical step beyond a CPL is what we will refer to as a running power law (RPL), a model in which the spectral index is allowed to exhibit a logarithmic frequency dependence, 
\begin{tcolorbox}
\begin{equation}
\label{eq:gammarun}
\gamma_{\rm run}\left(f\right) = \gamma + \beta\,\ln\left(\frac{f}{f_{\rm ref}}\right) 
\end{equation}
\end{tcolorbox}
As we will see shortly, the RPL model describes parabola-shaped GWB spectra, if plotted on doubly logarithmic axes, rendering it a natural generalization of the CPL model. While the RPL model has received only little attention in the PTA literature thus far [see \cite{Ben-Dayan:2023lwd} for a notable exception], similar constructions are well established in the literature on the cosmic microwave background (CMB). In their analysis of the primordial curvature power spectrum, the PLANCK Collaboration, e.g., uses their CMB data to constrain the ``running of the scalar spectral index'' and even the ``running of the running of the scalar spectral index''~\citep{Planck:2018vyg,Planck:2018jri}. The goal of the present paper is to introduce some of these ideas to the PTA community and initiate a systematic investigation of observational limits on the running of the spectral index in the PTA band, i.e, the parameter
\begin{equation}
\label{eq:beta}
\beta = \frac{d\gamma_{\rm run}\left(f\right)}{d \ln f} \,.
\end{equation}

The RPL model promises to serve as a better proxy for many GWB candidate models that have been proposed as a possible explanation for the signal in the PTA frequency band than the CPL model. On the astrophysical side, this is true because SMBHB models often predict a spectral turnover at low frequencies due to interactions with the circumbinary environment~\citep{Kocsis:2010xa,NANOGrav:2023hfp}, alongside a spectral break at high frequencies caused by the discrete nature of the SMBHB population~\citep{Sesana:2008mz,Agazie:2024jbf}. While other templates may be able to describe such features even better, the RPL model can at least roughly account for the presence of a spectral turnover or break, while the CPL model has no chance of doing so whatsoever. The real strength of the RPL model, however, lies in the fact that it can serve as a good or even very good approximation of many cosmological models. The spectral index of cosmological signals often varies slowly across several orders of magnitude in frequency space, which results in a mild running of the spectral index in the PTA frequency band and hence gives rise to an RPL-like spectral shape. We therefore argue that bounds on the three parameters of the RPL model\,---\,the amplitude $A$ at $f = f_{\rm ref}$, the spectral index $\gamma$ at $f = f_{\rm ref}$, and the running of the spectral index $\beta$\,---\,provide valuable information that can be used to constrain a large class of cosmological models. In this paper, we will fit the RPL model to the NG15 data to derive Bayesian limits on $A$, $\gamma$, $\beta$ of exactly this type. 

The rest of this paper is organized as follows. In the next section, we will properly define the RPL model and discuss some of its properties. In Sec.~\ref{sec:fit}, we will then perform a Bayesian fit of the RPL model to the NG15 data. This analysis will provide us with marginalized posterior distributions for the parameters $A$, $\gamma$, and $\beta$, which we will use to construct Bayesian $95\,\%$ credible intervals for all three parameters. Furthermore, we will carry out a Bayesian model comparison with the CPL model and determine the RPL-versus-CPL Bayes factor. In Sec.~\ref{sec:igw}, we will subsequently broaden the scope of our analysis and interpret the RPL model as a description of primordial GWs from cosmic inflation. This means we will extrapolate our results to frequencies above and below the PTA frequency band, which will allow us to combine our constraints on the RPL model with limits from big-bang nucleosynthesis (BBN), the CMB, and LIGO--Virgo--KAGRA (LVK). Section~\ref{sec:conclusions}, finally, contains our conclusion and a brief outlook on possible future applications of the RPL model. 


\section{Running power law} 
\label{sec:RPL}

\subsection{Spectral model}

The central observables in the context of PTA searches for GWs are the timing residuals $R_a$ for a set of galactic millisecond pulsars. The imprint of a stochastic GWB on these timing residuals manifests itself in an extra contribution to the timing-residual cross-power spectrum for pairs of pulsars $a$ and $b$~\citep{NANOGrav:2023icp},
\begin{equation}
\label{eq:Sab}
S_{ab}^{\rm GW}\left(f\right) = \Gamma_{ab}\, \frac{S_h\left(f\right)}{6\pi^2 f^2} \,.
\end{equation}
Here, $\Gamma_{ab}$ denotes the HD cross-correlation coefficients,
\begin{equation}
\Gamma_{ab}\left(\xi_{ab}\right) = \left(1+\delta_{ab}\right) \left[\frac{3}{2}\,x_{ab} \ln x_{ab} -\frac{x_{ab}}{4} + \frac{1}{2} \right] \,, 
\end{equation}
with $x_{ab} = \sfrac{1}{2}\left(1-\cos\xi_{ab}\right)$ and $\xi_{ab}$ being the angular separation between pulsar $a$ and pulsar $b$ in the sky. $S_h$ is the GW strain power spectrum, which is closely related to the characteristic GW strain amplitude,
\begin{equation}
h_c\left(f\right) = \sqrt{2f\,S_h\left(f\right)} \,,
\end{equation}
as well as to the GW energy density spectrum in units of the critical energy density of the present Universe,
\begin{equation}
\Omega_{\rm GW}\left(f\right) = \frac{1}{\rho_{\rm crit}}\,\frac{d\rho_{\rm GW}\left(f\right)}{d \ln f} = \frac{4\pi^2}{3H_0^2}\,f^3\,S_h\left(f\right) \,,
\end{equation}
where $H_0 = 100\,h\,\textrm{km}/\textrm{s}/\textrm{Mpc}$ is the Hubble constant and $h \sim 0.7$~\citep{Kamionkowski:2022pkx}. Below, we will always work with $h^2 \Omega_{\rm GW}$, which, unlike $\Omega_{\rm GW}$, is independent of the precise value of the Hubble constant.

The CPL model starts from a power-law ansatz for $h_c$, 
\begin{equation}
h_c\left(f\right) \:\:\overset{\textrm{CPL}}{=}\:\: A\left(\frac{f}{f_{\rm ref}}\right)^\alpha  \,,
\end{equation}
which immediately translates to a power law for $S_{ab}^{\rm GW}$,
\begin{equation}
\label{eq:SabGWCPL}
S_{ab}^{\rm GW}\left(f\right) \:\:\overset{\textrm{CPL}}{=}\:\: \Gamma_{ab}\, \frac{A^2}{12\pi^2 f_{\rm ref}^3} \left(\frac{f}{f_{\rm ref}}\right)^{-\gamma} \,,
\end{equation}
with spectral index $\gamma = 3-2\alpha$, and similarly for $\Omega_{\rm GW}$,
\begin{equation}
\Omega_{\rm GW}\left(f\right) \:\:\overset{\textrm{CPL}}{=}\:\: \frac{2\pi^2}{3H_0^2}\,A^2 f_{\rm ref}^2 \left(\frac{f}{f_{\rm ref}}\right)^n \,,
\end{equation}
with spectral index $n = 5 - \gamma = 2\alpha + 2$.
The relation in Eq.~\eqref{eq:SabGWCPL} introduces the spectral index $\gamma$ that we already referred to in Sec.~\ref{sec:introduction}. From this relation, $-\gamma$ can be identified with the coefficient of $x = \ln(f/f_{\rm ref})$ after taking the logarithm of both sides of the equation,
\begin{equation}
\label{eq:gammadef1}
\ln S_{ab}^{\rm GW}\left(x\right) \:\:\overset{\textrm{CPL}}{=}\:\: \ln\left(\Gamma_{ab}\, \frac{A^2}{12\pi^2 f_{\rm ref}^3}\right) - \gamma \,x \,,
\end{equation}
which can also be written as 
\begin{equation}
\label{eq:gammadef15}
\gamma \:\:\overset{\textrm{CPL}}{=}\:\: - \frac{\ln S_{ab}^{\rm GW}\left(x\right) - \ln S_{ab}^{\rm GW}\left(x=0\right)}{x} \,.
\end{equation}
Alternatively, $-\gamma$ can be recovered from the derivative of the log of the timing-residual cross-power spectrum,
\begin{equation}
\label{eq:gammadef2}
\gamma \:\:\overset{\textrm{CPL}}{=}\:\: - \frac{d\ln S_{ab}^{\rm GW}}{dx} \,.
\end{equation}
Both approaches yield, of course, the same result: the coefficient $\gamma$ in front of the term linear in $x$ in Eq.~\eqref{eq:gammadef1} coincides with $\gamma$ in Eq.~\eqref{eq:gammadef2}. To see this, simply differentiate both sides of Eq.~\eqref{eq:gammadef1} with respect to $x$.

We shall now generalize the CPL model and allow for a running of the spectral index $\gamma$. To this end, we shall replace $\gamma$ on the left-hand side of Eq.~\eqref{eq:gammadef2} by our ansatz for the running spectral index $\gamma_{\rm run}$ in Eq.~\eqref{eq:gammarun}, which yields a first-order differential equation for $S_{ab}^{\rm GW}$,
\begin{equation}
\label{eq:DGL}
\gamma_{\rm run}\left(x\right) = \gamma + \beta\,x \:\:\overset{\textrm{RPL}}{=}\:\: - \frac{d\ln S_{ab}^{\rm GW}}{dx} \,.
\end{equation}
Then, imposing the same boundary condition as before,
\begin{equation}
S_{ab}^{\rm GW}\left(x=0\right) \:\:\overset{\textrm{RPL}}{=}\:\: \Gamma_{ab}\, \frac{A^2}{12\pi^2 f_{\rm ref}^3} \,,
\end{equation}
we can immediately write down the solution of Eq.~\eqref{eq:DGL},
\begin{tcolorbox}
\vspace{-0.45cm}
\begin{equation}
\label{eq:SabGWRPL}
S_{ab}^{\rm GW}\left(f\right) \:\:\overset{\textrm{RPL}}{=}\:\: \Gamma_{ab}\, \frac{A^2}{12\pi^2 f_{\rm ref}^3} \left(\frac{f}{f_{\rm ref}}\right)^{-\tilde{\gamma}_{\rm run}\left(f\right)}
\end{equation}
\end{tcolorbox}

\noindent
with the index $\tilde{\gamma}_{\rm run}$ in the exponent being given as
\begin{equation}
\tilde{\gamma}_{\rm run}\left(f\right) = \gamma + \frac{1}{2}\,\beta\,\ln\left(\frac{f}{f_{\rm ref}}\right) \,.
\end{equation}

Clearly, the two possible definitions of the spectral index that we encountered in Eqs.~\eqref{eq:gammadef15} and \eqref{eq:gammadef2} now no longer agree, $\gamma_{\rm run} \neq \tilde{\gamma}_{\rm run}$. In this paper, we will refer to $\gamma_{\rm run}$ as \textit{the} running spectral index and $\tilde{\gamma}_{\rm run}$ as the \textit{naive} running spectral index, because of the former's more useful geometric interpretation: in a plot with log--log axes, $-\gamma_{\rm run}$ directly measures the instantaneous slope of $S_{ab}^{\rm GW}(x)$ at $x$, while $-\tilde{\gamma}_{\rm run}$ measures the slope of the straight line connecting $S_{ab}^{\rm GW}(x)$ to $S_{ab}^{\rm GW}(x=0)$, which is of less interest in the RPL model. Both versions of the spectral index are related to each other via
\begin{equation}
\gamma_{\rm run}\left(x\right) = \frac{d}{dx}\left(\tilde{\gamma}_{\rm run}\left(x\right)x\right) \,.
\end{equation}
This relation is true in the RPL model;  but it also holds in more general models for analogously defined indices $\gamma_{\rm run}$ and $\tilde{\gamma}_{\rm run}$ with arbitrary frequency dependence.

The relation in Eq.~\eqref{eq:SabGWRPL} defines the RPL model at the level of the timing-residual cross-power spectrum. Similarly to the CPL model, we can express the information on the GWB also in terms of the characteristic GW strain amplitude, 
\begin{equation}
h_c\left(f\right) \:\:\overset{\textrm{RPL}}{=}\:\: A\left(\frac{f}{f_{\rm ref}}\right)^{\tilde{\alpha}_{\rm run}\left(f\right)}  \,,
\end{equation}
with naive running spectral index 
\begin{equation}
\tilde{\alpha}_{\rm run}\left(f\right) = \frac{3}{2} - \frac{1}{2}\left[\gamma + \frac{1}{2}\,\beta\,\ln\left(\frac{f}{f_{\rm ref}}\right)\right] \,,
\end{equation}
as well as in terms of the GW energy density spectrum,
\begin{equation}
\label{eq:OmegaRPL}
\Omega_{\rm GW}\left(f\right) \:\:\overset{\textrm{RPL}}{=}\:\: \frac{2\pi^2}{3H_0^2}\,A^2 f_{\rm ref}^2 \left(\frac{f}{f_{\rm ref}}\right)^{\tilde{n}_{\rm run}\left(f\right)} \,,
\end{equation}
with naive running spectral index 
\begin{equation}
\tilde{n}_{\rm run}\left(f\right) = 5 - \left[\gamma + \frac{1}{2}\,\beta\,\ln\left(\frac{f}{f_{\rm ref}}\right)\right] \,.
\end{equation}

These results illustrate that the RPL model does indeed give rise to parabola-shaped GWB spectra, i.e, $\ln h^2\Omega_{\rm GW}$ is a second-order polynomial in $\ln f$,
\begin{equation}
\ln h^2\Omega_{\rm GW}\left(x\right) = c_0 + c_1\,x + \frac{1}{2}\,c_2\,x^2 \,,
\end{equation}
where $x = \ln(f/f_{\rm ref})$ as before and with coefficients
\begin{align}
c_0 & = \left.\ln h^2\Omega_{\rm GW}\right|_{x=0} = \ln \left(\frac{2\pi^2}{3H_0^2}\,h^2A^2 f_{\rm ref}^2\right) \,, \\
c_1 & = \left.\frac{d \ln h^2\Omega_{\rm GW}}{d x}\right|_{x=0} = 5 - \gamma \,, \\
\label{eq:c2}
c_2 & = \left.\frac{d^2 \ln h^2\Omega_{\rm GW}}{d x^2}\right|_{x=0} = -\beta \,.
\end{align}
The relation in Eq.~\eqref{eq:c2} illustrates that, up to a negative sign, the new parameter $\beta$ can be interpreted as the curvature of the GWB spectrum on log--log axes, which agrees with the convention in \cite{Ben-Dayan:2023lwd}. In the RPL model, this curvature is, in fact, constant, such that we do not have to restrict ourselves to $x=0$,
\begin{equation}
\beta = - \frac{d^2 \ln h^2\Omega_{\rm GW}}{d x^2} = \textrm{const} \,.
\end{equation}
Here, the relative sign simply follows from the sign convention in Eq.~\eqref{eq:SabGWRPL}, according to which $S_{ab}^{\rm GW} \propto f^{-\tilde{\gamma}_{\rm run}}$. 


\begin{table*}
\renewcommand{\arraystretch}{1.2}

\caption{Prior probability density distributions for the free parameters in our fit of the RPL model to the NG15 data.}
\label{tab:priors}

\begin{center}
\begin{tabular}{llll}
\toprule
\textbf{Parameter} & \textbf{Description} & \textbf{Prior} & \textbf{Comments}                          \\ [4pt] \hline
\multicolumn{4}{c}{\textbf{Pulsar-intrinsic red noise}}                                                 \\
$A_{\rm red}$      &amplitude                      & log-uniform $[-18,-12]$ & one parameter per pulsar \\
$\gamma_{\rm red}$ & spectral index            & uniform $[0,10]$         & one parameter per pulsar \\ [4pt] \hline
\multicolumn{4}{c}{\textbf{GWB in the RPL model}} \\ 
$A\left(\sfrac{1}{10\,\textrm{yr}}\right)$      & amplitude                     & log-uniform $[-18,-12]$ & one parameter per PTA    \\
$\gamma\left(\sfrac{1}{10\,\textrm{yr}}\right)$ & spectral index                & uniform $[0,10]$         & one parameter per PTA    \\
$\beta$                                          & running of the spectral index & uniform $[-2,4]$      & one parameter per PTA    \\
\bottomrule 
\end{tabular}
\end{center}

\end{table*}


\subsection{Reference frequency}

Both the CPL model and the RPL model require one to specify a reference frequency $f_{\rm ref}$. This frequency, however, only serves as an auxiliary quantity and does not represent an independent physical parameter. In the CPL model, $f_{\rm ref}$ determines the physical meaning of the amplitude $A$: the value of the characteristic GW strain amplitude $h_c$ at $f = f_{\rm ref}$. Meanwhile, $\gamma$ is a constant and hence independent of the choice of $f_{\rm ref}$ in the CPL model. It is thus straightforward to translate any pair of values $\left(A,\gamma\right)$ from one choice of $f_{\rm ref}$ to another,
\begin{equation}
\label{eq:frefCPL}
\begin{pmatrix} \ln A' \\ \gamma' \end{pmatrix}  \:\:\overset{\textrm{CPL}}{=}\:\: \begin{pmatrix} 1 & -\frac{R}{2} \\ 0 & 1\end{pmatrix} \begin{pmatrix} \ln A \\ \gamma \end{pmatrix} + \begin{pmatrix} \frac{3R}{2} \\ 0 \end{pmatrix} \,,
\end{equation}
where $R = \ln(f_{\rm ref}'/f_{\rm ref}^{\vphantom{\prime}})$. The affine relation in Eq.~\eqref{eq:frefCPL} expresses the simple idea that $\gamma' = \gamma$ and 
\begin{equation}
A' \:\:\overset{\textrm{CPL}}{=}\:\: A\left(\frac{f_{\rm ref}'}{f_{\rm ref}}\right)^\alpha \,,
\end{equation}
for any two choices of the reference frequency, $f_{\rm ref}$ and $f_{\rm ref}'$, in the CPL model.

The situation in the RPL model is analogous. The choice of $f_{\rm ref}$ determines the physical meaning of $A$ and $\gamma$: $A$ is again the value of $h_c$ at $f = f_{\rm ref}$, and $-\gamma$ is the spectral index of $S_{ab}^{\rm GW}$ at $f = f_{\rm ref}$. Meanwhile, the new parameter $\beta$ is a constant and independent of $f_{\rm ref}$. Requiring the actual values of $S_{ab}^{\rm GW}$ to be invariant under a change of the reference frequency, $f_{\rm ref} \rightarrow f_{\rm ref}'$, we now obtain the following affine transformation,
\begin{equation}
\label{eq:frefRPL}
\begin{pmatrix} \ln A' \\ \gamma' \\ \beta' \end{pmatrix}  \:\:\overset{\textrm{RPL}}{=}\:\ \begin{pmatrix} 1 & -\frac{R}{2} & -\frac{R^2}{4} \\ 0 & 1 & R \\ 0 & 0 & 1 \end{pmatrix} \begin{pmatrix} \ln A \\ \gamma \\ \beta \end{pmatrix} + \begin{pmatrix} \frac{3R}{2} \\ 0 \\ 0 \end{pmatrix} \,,
\end{equation}
where the behavior of $A$ reflects the idea that
\begin{equation}
A' \:\:\overset{\textrm{RPL}}{=}\:\: A\left(\frac{f_{\rm ref}'}{f_{\rm ref}}\right)^{\tilde{\alpha}\left(f_{\rm ref}'\right)} \,,
\end{equation}
while the transformation behavior of $\gamma$ amounts to
\begin{equation}
\gamma' \:\:\overset{\textrm{RPL}}{=}\:\: \gamma_{\rm run}\left(f_{\rm ref}'\right) \,,
\end{equation}
in direct analogy to $\gamma = \gamma_{\rm run}(f_{\rm ref})$. Meanwhile, the running of the spectral index remains constant, $\beta'=\beta$.

Below, we will set $f_{\rm ref} = \sfrac{1}{10\,\textrm{yr}} \simeq 3.17\,\textrm{nHz}$, which falls into the range of frequencies where NANOGrav's sensitivity to a GWB signal is maximal [see the NANOGrav sensitivity curve in \cite{NANOGrav:2023ctt}]. This choice of reference frequency has the advantage that it helps to minimize the covariance among the parameters $A$, $\gamma$, and $\beta$ in our Bayesian fit analysis. On the other hand, our choice of $f_{\rm ref}$ is still arbitrary; and it is an easy exercise to convert our result for the 3D posterior density of $A$, $\gamma$, and $\beta$ that we will discuss in the next section to the posterior density at a different reference frequency.

Indeed, this conversion is straightforward at the level of the Monte Carlo Markov chain (MCMC) that we obtain from our Bayesian fit analysis: to convert from $f_{\rm ref}$ to $f_{\rm ref}'$, one simply has to apply Eq.~\eqref{eq:frefRPL} to the MCMC on a sample-by-sample basis. In this way, one automatically accounts for the transformation behavior of the volume element in the 3D parameter space, which is crucial in order to correctly describe the transformation behavior of the posterior \textit{density}. At the same time, we caution that Eq.~\eqref{eq:frefRPL} cannot be used to convert marginalized 2D or 1D posterior densities from one choice of reference frequency to another, because marginalization and the transformation in Eq.~\eqref{eq:frefRPL} do not commute. If one is interested in marginalized 2D or 1D posterior densities at a different reference frequency, one must first transform the full 3D posterior density according to Eq.~\eqref{eq:frefRPL} and then marginalize, not vice versa. 


\section{Bayesian fit to the NG15 data}
\label{sec:fit}

\subsection{Tools and methods}
\label{subsec:tools}

We now turn to our Bayesian fit of the RPL model to the NG15 data. Our analysis closely follows~\cite{NANOGrav:2023hvm}, where more details on the underlying formalism and conventions can be found. Similarly to~\cite{NANOGrav:2023hvm}, we use our software package \texttt{PTArcade}~\citep{Mitridate:2023oar}, a wrapper for \texttt{ENTERPRISE}~\citep{2019ascl.soft12015E} and \texttt{ENTERPRISE\_EXTENSIONS}~\citep{enterprise}, to implement the RPL model and fit it to the NG15 data. \texttt{PTArcade} can be run in two different modes: ``enterprise'' and ``ceffyl'' \citep{Lamb:2023jls}. We run it in ``enterprise'' mode, which means that we carry out a full Bayesian MCMC analysis in the time domain. 

The different contributions to the NG15 timing residuals are treated in the same way as in~\cite{NANOGrav:2023hvm}: (i) all white-noise parameters are kept fixed at the maximum a posteriori (MAP) values recovered from single-pulsar analyses~\citep{NANOGrav:2023ctt}, (ii) linear offsets in the timing-ephemerides parameters are marginalized over, and (iii) pulsar-intrinsic red noise as well as the GWB signal are modeled using a discrete frequency basis $f_i = i/T$ (where $T\simeq 16.03\,\textrm{yr}$ is the total extent of the data set) in Fourier space. In addition to the linear offsets in the timing model, we also marginalize over the coefficients of this Fourier series, which results in the standard marginalized PTA likelihood~\citep{vanHaasteren:2012hj,Lentati:2012xb}. Meanwhile, we model pulse dispersion, again following~\cite{NANOGrav:2023hvm}, as a piecewise constant with the inclusion of DMX parameters~\citep{NANOGrav:2015qfw,Jones:2016fkk}.


\begin{table*}
\renewcommand{\arraystretch}{1.2}

\caption{Point and interval estimates for the parameters of the RPL model. Column 2 states the maximum a posteriori (MAP) value for each parameter based on the full 3D posterior density. Columns 3 to 5 state the MAP values and highest-posterior-density intervals at the $68\,\%$ and $95\,\%$ credible level based on the three marginalized 1D posterior densities in Fig.~\ref{fig:corner}.}
\label{tab:posteriors}

\begin{center}
\begin{tabular}{lrrrr}
\toprule
\textbf{Parameter}                                    & \textbf{3D MAP value}    & \textbf{1D MAP value}    & \textbf{\boldmath{$68\,\%$} credible interval} & \textbf{\boldmath{$95\,\%$} credible interval} \\ [4pt] \hline
$\log_{10} A\left(\sfrac{1}{10\,\textrm{yr}}\right)$ & $-14.09$ & $-14.09$ & $[-14.17,-14.00]$ & $[-14.25,-13.91]$ \\
$\gamma\left(\sfrac{1}{10\,\textrm{yr}}\right)$      & $2.68$ & $2.60$   & $[1.81,3.35]$     & $[0.98,4.05]$    \\
$\beta$                                               & $0.74$ & $0.92$   & $[0.01,1.90]$    & $[-0.80,2.96]$    \\
\bottomrule 
\end{tabular}
\end{center}

\end{table*}


The only remaining free parameters in the marginalized likelihood are the parameters entering the covariance matrix of the time-correlated stochastic processes, i.e., pulsar-intrinsic red noise and the GWB signal. For each individual pulsar, we model red noise in terms of a power law with amplitude $A_{\rm red}$ and spectral index $\gamma_{\rm red}$, spanning from $f_1 = 1/T$ to $f_{30} = 30/T$. Meanwhile, the GWB signal is modeled in terms of the RPL expression for the timing-residual cross-power spectrum in Eq.~\eqref{eq:SabGWRPL}. As the GWB signal in the NG15 data mostly appears at lower frequencies, $f \lesssim f_{14} = 14/T$, we include the GWB contribution to the timing residuals only in the first 14 frequency bins. In total, this leaves us with $137$ free parameters: $A_{\rm red}$ and $\gamma_{\rm red}$ for 67 pulsars, plus the three free parameters of the RPL model, $A$, $\gamma$, $\beta$. Each point in the space spanned by these 137 parameters defines a statistical ensemble of possible realizations of pulsar-intrinsic red noise and the GWB that all derive from the same covariance matrix but which differ in terms of their explicit coefficients in Fourier space. The marginalized PTA likelihood no longer depends on these coefficients, but serves as a likelihood on the 137-dimensional space of parameters (sometimes also called ``hyperparameters'') that control the power spectra of pulsar-intrinsic red noise and the GWB. In our Bayesian analysis, we sample from the posterior density of our 137 model parameters using MCMC techniques~\citep{justin_ellis_2017_1037579} applied to the marginalized PTA likelihood and the prior densities summarized in Table~\ref{tab:priors}. Our priors are agnostic about the source of the GWB signal and large enough so that they can accommodate the entire 3D $95\,\%$ posterior volume of the RPL model. Our final MCMC chain encompasses nearly three million samples and is composed of 20 independent subchains.

The CPL model can be fitted to the NG15 data in the same way as the RPL model, the only difference being that $\beta$ needs to be set to $\beta = 0$ at all times. We make use of this fact and employ product-space methods~\citep{10.2307/2346151,10.2307/1391010,10.1093/mnras/stv2217} to fit both models simultaneously and thus determine the Bayes factor $\mathcal{B}$ for the RPL-versus-CPL model comparison. In fact, we use statistical bootstrapping methods to determine a mean value and a standard deviation for the Bayes factor~\citep{Efron:1986hys} [see the discussion in~\cite{NANOGrav:2023hvm} for more details]. 

Before we turn to the discussion of our results, let us also briefly compare our analysis to one in \cite{Ben-Dayan:2023lwd}. While we work with the marginalized PTA likelihood for the NG15 timing residuals, \cite{Ben-Dayan:2023lwd} start from the 2D posterior density for $A$ and $\gamma$ in the CPL model. They extract this posterior density from \cite{NANOGrav:2023gor}, reinterpret it as a likelihood, and then use this likelihood to fit various GWB models, including a modification of the RPL model that also accounts for the dynamics of reheating after inflation. That is, they refit an RPL-like model to the 2D posterior density of the CPL model, which provides them with tight bounds on $c_2 = -\beta$ in Eq.~\eqref{eq:c2}, $\left|c_2\right| \lesssim 0.12$. As we will now discuss, our analysis results in less tight bounds on $\beta$, as we do not start from the 2D posterior density of the CPL model, but allow our MCMC sampler to explore the whole 3D parameter space of the RPL model subject to the priors in Table~\ref{tab:priors}. 


\subsection{Results}

\begin{figure}[t]
\begin{center}
\includegraphics[width=0.47\textwidth]{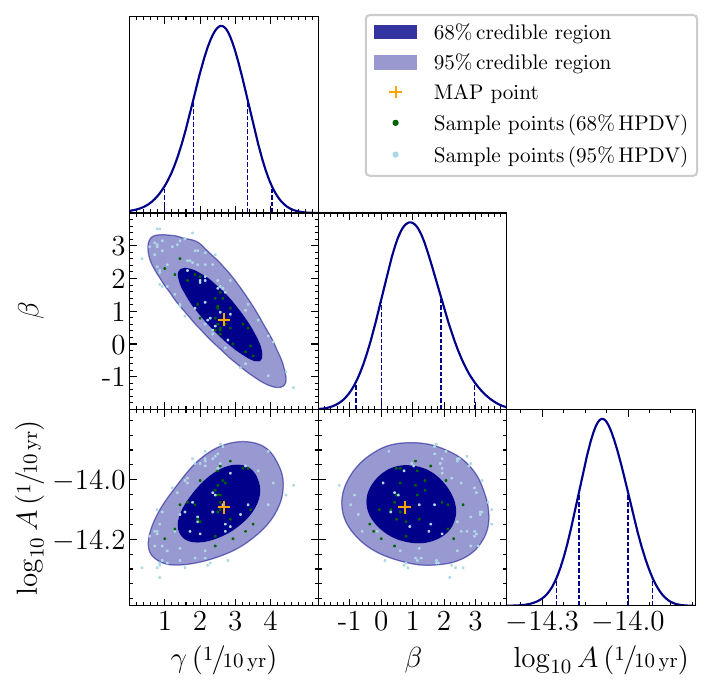}
\end{center}
\caption{Corner plot of the 3D parameter space of the RPL model. The blue-shaded regions in the off-diagonal plots and the solid blue curves in the diagonal plots show the marginalized 2D and 1D posterior densities for the RPL model parameters, respectively. The dark-blue regions mark $68\,\%$ credible regions, the light-blue regions mark $95\,\%$ credible regions, and the dashed vertical lines mark $68\,\%$ and $95\,\%$ credible intervals. The cyan- and green-colored points represent random samples from the 68\,\% and 95\,\% highest-posterior-density volumes (HPDVs) in the 3D parameter space, respectively. The orange cross marks the point where the 3D posterior density reaches its maximum (see the second column in Table~\ref{tab:posteriors}). The GWB spectra corresponding to the colored points are shown in identical colors in Fig.~\ref{fig:spectrum}.}
\label{fig:corner}
\end{figure}

The main result of our Bayesian fit analysis is the 3D posterior density for the RPL parameters $A$, $\gamma$, and $\beta$. The marginalized 2D and 1D posterior densities that can be derived from this 3D density are shown in the corner plot in Fig.~\ref{fig:corner}. In one, two, and three dimensions, we rely on kernel density estimation (KDE) [specifically, the \texttt{GetDist} package~\cite{Lewis:2019xzd}] in order to construct smooth densities from discrete sets of MCMC samples.

The KDE approximation of the 3D posterior density allows us to determine the MAP point in parameter space, i.e., the point of highest 3D posterior density. We list the coordinates of this point in the second column of Table~\ref{tab:posteriors} and mark its position in Fig.~\ref{fig:corner} with an orange-colored cross. Similarly, we can derive point and interval estimates from the three marginalized 1D posterior densities shown in Fig.~\ref{fig:corner}. Specifically, we determine the MAP values of $A$, $\gamma$, and $\beta$ according to their respective 1D densities (see the third column in Table~\ref{tab:posteriors}) and the corresponding $68\,\%$ and $95\,\%$ credible intervals (see the fourth and fifth columns in Table~\ref{tab:posteriors}). Here, the credible intervals are defined as highest-posterior-density intervals (HPDIs), i.e., we integrate the 1D densities over regions of highest posterior density until the integral returns an integrated probability of $68\,\%$ or $95\,\%$.

Based on the values listed in Table~\ref{tab:posteriors}, we are unable to claim evidence for nonzero running of the spectral index in the PTA frequency band; our Bayesian interval estimates for $\beta$ are perfectly consistent with the assumption of no running, $\beta = 0$. Conversely, there is nothing wrong with assuming nonzero running. In fact, our results indicate that $\beta$ can easily be of $\mathcal{O}(1)$. The $68\,\%$ credible interval for $\beta$ even barely includes $\beta = 0$, and the 3D and 1D MAP values of $\beta$ are clearly nonzero. We therefore conclude that, while the NG15 data does not yet suffice to claim the discovery of nonzero $\beta$, future PTA data sets may well enable such a measurement. 

\begin{figure}
\begin{center}
\includegraphics[width=0.47\textwidth]{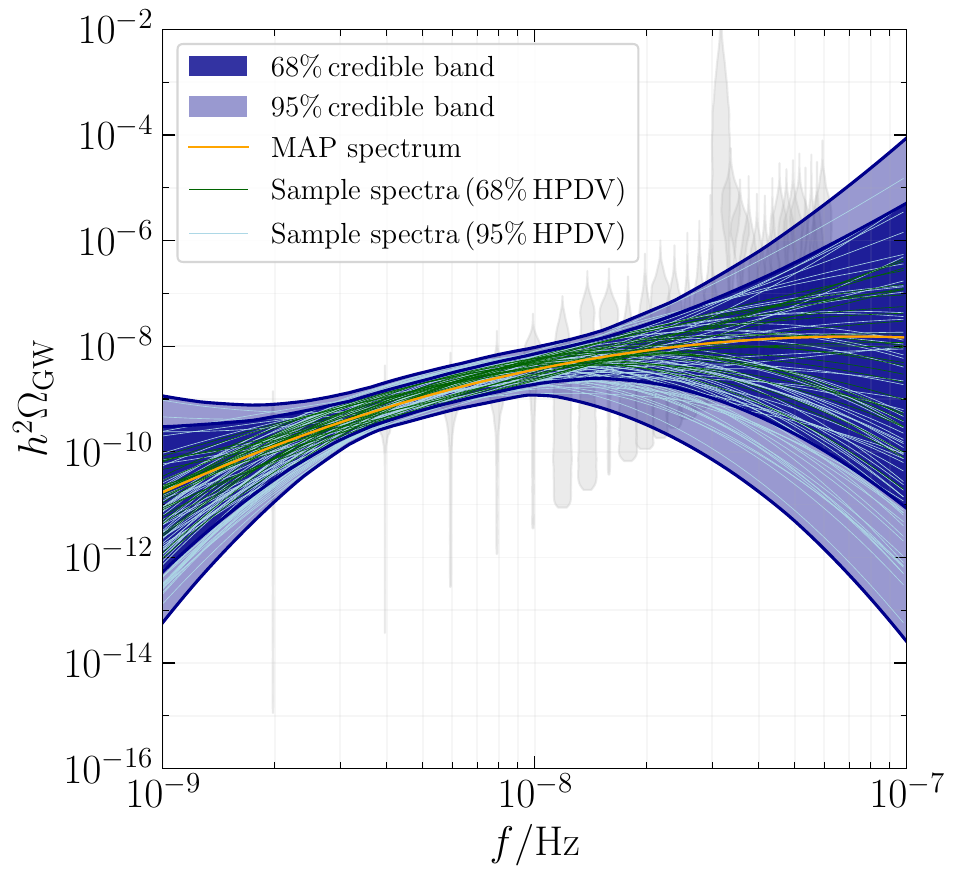}
\end{center}
\caption{GWB spectra predicted by the RPL model. The blue-shaded bands encompass all spectra that are associated with points inside the 68\,\% and 95\,\% highest-posterior-density volumes (HPDVs) in the 3D parameter space. The cyan-, green, and orange-colored spectra belong to the parameter points shown in Fig.~\ref{fig:corner}. The gray ``violins'' in the background represent the posterior densities for the 30 values of the GW energy density spectrum, $\{h^2\Omega_{\rm GW}(f_i)\}$ with $f_i = i/T$ and $i = 1\cdots 30$, in the free spectral model.}
\label{fig:spectrum}
\end{figure}

Besides parameter inference, we are also interested in exploring the implications of our analysis for the GWB spectrum. We do this in two different ways: 

\smallskip\noindent
\textbf{(1) Bayesian credible bands:} In Fig.~\ref{fig:spectrum}, we show what we shall refer to as $68\,\%$ and $95\,\%$ credible bands for the GWB spectrum predicted by the RPL model. These bands are based on the $68\,\%$ and $95\,\%$ highest-posterior-density volumes (HPDVs) in the 3D parameter space. In fact, they can be regarded as the projection of these HPDVs onto the space of possible GWB spectra. In order to construct the HPDVs in the 3D parameter space, we proceed in the same way as for the HPDIs in the marginalized 1D posterior densities: we integrate the KDE approximation of the 3D density over regions of highest posterior density until the integral returns an integrated probability of $68\,\%$ or $95\,\%$. In the next step, we then take all points inside these HPDVs, compute the GWB spectra that they predict, and draw all these spectra in a plot of $h^2\Omega_{\rm GW}$ as a function of $f$. This procedure results in two families of GWB spectra (spectra belonging to points in the $68\,\%$ HPDV and spectra belonging to points in the $95\,\%$ HPDVs). The envelopes of these two families of GWB spectra define the $68\,\%$ and $95\,\%$ credible bands shown in Fig.~\ref{fig:spectrum}. In order to illustrate the algorithm behind this construction, we show a handful of sample points in the $68\,\%$ or $95\,\%$ HPDVs in Fig.~\ref{fig:corner} and their associated GWB spectra in identical colors in Fig.~\ref{fig:spectrum}. In the limit of a large number of samples in Fig.~\ref{fig:corner}, the sample spectra in Fig.~\ref{fig:spectrum} begin to shape out the $68\,\%$ and $95\,\%$ credible bands. On top, we show the position of the 3D MAP point in Fig.~\ref{fig:corner} and the GWB spectrum that it predicts in Fig.~\ref{fig:spectrum}.


\begin{figure}
\begin{center}
\includegraphics[width=0.47\textwidth]{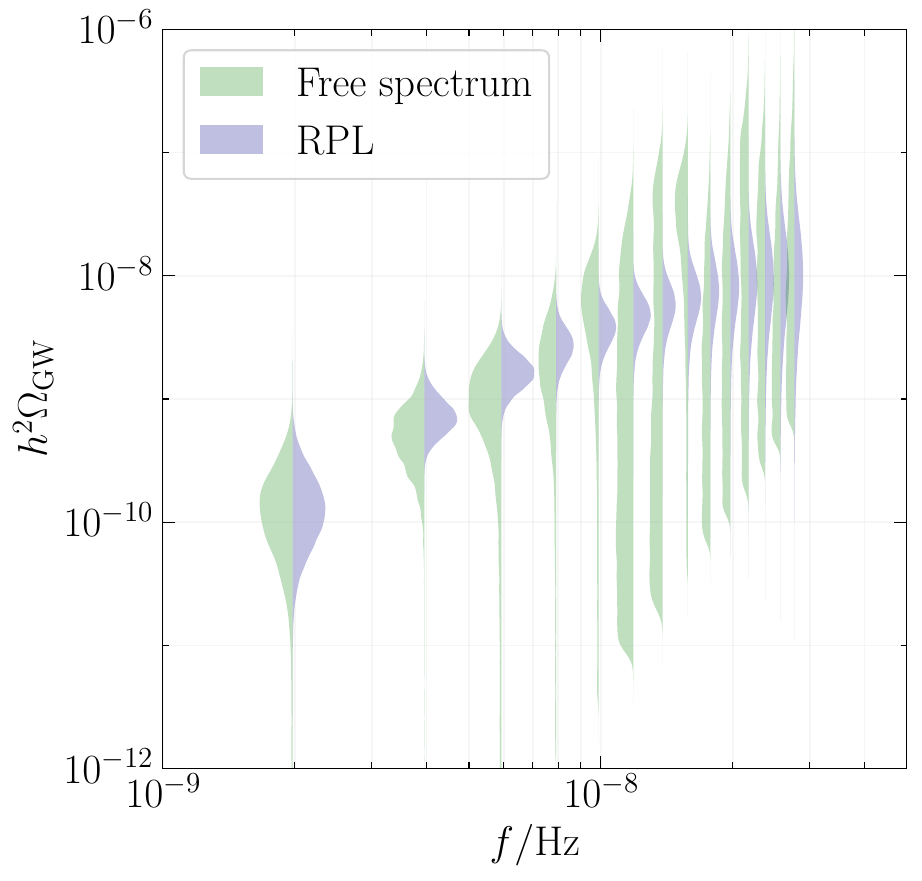}
\end{center}
\caption{Bayesian periodograms for the free spectral model (green ``violins'') and the RPL model (purple ``violins''). Each ``half-violin'' corresponds to a Bayesian posterior probability density for the respective value of $h^2\Omega_{\rm GW}(f_i)$.}
\label{fig:periodogram}
\end{figure}

\smallskip\noindent
\textbf{(2) Bayesian periodogram:} In Fig.~\ref{fig:periodogram}, we present the Bayesian periodogram for the RPL model and compare it to the Bayesian periodogram for the free spectral model. Both periodograms display the posterior densities for the 14 values of the GW energy density spectrum, $\{h^2\Omega_{\rm GW}(f_i)\}$, at $f_i = i/T$ with $i = 1\cdots 14$. Given a large number of MCMC samples, these posterior densities can be simply constructed from histograms of $h^2\Omega_{\rm GW}(f_i)$ values (i.e., one histogram at each frequency $f_i$), in combination with a KDE approximation. Specifically, in order to create the periodograms in Fig.~\ref{fig:periodogram}, we used tools included in NANOGrav's \texttt{holodeck} software package~\citep{NANOGrav:2023hfp}. The periodogram of the free spectral model corresponds to what is better known as the NG15 ``violins''; these ``violins'' (in their complete, two-sided form) are also shown in gray in Fig.~\ref{fig:spectrum}. Similarly, the periodogram of the RPL model may be referred to as the ``RPL violins''. Based on the periodogram of the RPL model, one could in principle construct what we referred to as the ``median GWB spectrum'' in~\cite{NANOGrav:2023hvm}: a curve that connects the median of the first RPL violin with the median of the second RPL violin, and so on. However, a potentially misleading issue related to the concept of median GWB spectra is that they almost never coincide with an individual GWB spectrum at a certain point in the model parameter space. In the case of the RPL model, this means that its median GWB spectrum would not exactly correspond to a parabola, even though every single individual GWB spectrum predicted by the RPL model is parabola-shaped. We therefore refrain from showing the median GWB spectrum of the RPL model and only present its periodogram, which in any case contains more information than just the median GWB spectrum by itself. Meanwhile, explicit GWB spectra that are predicted by the RPL model are shown in Fig.~\ref{fig:spectrum}.

\smallskip
The plots in Figs.~\ref{fig:spectrum} and \ref{fig:periodogram} contain complementary information. To see this, first note that the Bayesian periodogram in Fig.~\ref{fig:periodogram} results in $68\,\%$ and $95\,\%$ credible intervals for the values of $h^2\Omega_{\rm GW}(f_i)$ in each frequency bin $f_i$. These intervals, however, do not exactly coincide with the extent of the $68\,\%$ and $95\,\%$ credible bands in Fig.~\ref{fig:spectrum}. The $95\,\%$ band in Figs.~\ref{fig:spectrum}, e.g., is constructed from the $95\,\%$ of all MCMC samples that lie closest together in the parameter region of highest 3D posterior density. The $95\,\%$ intervals in Figs.~\ref{fig:periodogram}, on the other hand, belong to the $95\,\%$ of all MCMC samples whose values of $h^2\Omega_{\rm GW}(f_i)$ lie closest together (but whose parameter points may not necessarily lie close together). This means that, in both cases, $95\,\%$ of all MCMC samples are used\,---\,but not exactly the same selection. 

As a consequence, the plots in Figs.~\ref{fig:spectrum} and \ref{fig:periodogram} provide answers to two slightly different questions. The Bayesian credible bands in Fig.~\ref{fig:spectrum} better reflect the perspective of a model-building theorist who works under the assumption that the GWB signal is caused by a physical mechanism that indeed results in an RPL spectrum and that determines the true, physical values of $A$, $\gamma$, and $\beta$. If these physical values should correspond to the 3D MAP point in Fig.~\ref{fig:corner}, the MAP spectrum in Fig.~\ref{fig:spectrum} will be realized; if the physical values should slightly deviate from the 3D MAP point, the spectrum will slightly deviate from the MAP spectrum, and so on. In this sense, the credible bands in Fig.~\ref{fig:spectrum} tell us in which range we should expect the true spectrum to fall if we believe that the underlying mechanism singles out a specific region of parameter space. The Bayesian periodogram in Fig.~\ref{fig:periodogram}, on the other hand, remains ignorant towards the physical meaning of $A$, $\gamma$, and $\beta$. It better reflects the perspective of a theory-agnostic experimentalist who is primarily interested in the spectral shape of the GWB signal. From the periodogram, the experimentalist can read off in what proportion certain values of $h^2\Omega_{\rm GW}(f_i)$ are realized across all MCMC samples, independent of the precise values of $A$, $\gamma$, and $\beta$ that are needed to obtain these values of $h^2\Omega_{\rm GW}(f_i)$. In short, the credible bands in Fig.~\ref{fig:spectrum} group together our MCMC samples according to their $A$, $\gamma$, and $\beta$ values; the periodogram in Fig.~\ref{fig:periodogram} groups together our MCMC samples according to the shape of the GWB spectrum.


The main message from both Fig.~\ref{fig:spectrum} and Fig.~\ref{fig:periodogram} is that the RPL model provides a good fit of the NG15 data that is consistent with the free spectral reconstruction. Moreover, both figures yield an impression of how the ``violins'' of the free spectral reconstruction are probably going to evolve with more data in the future, if the GWB spectrum should indeed have an RPL-like shape.

We conclude our discussion by quoting the Bayes factor for the RPL-versus-CPL model comparison that we obtain from the hypermodel run described in Sec.~\ref{subsec:tools},
\begin{tcolorbox}
\begin{equation}
\mathcal{B}\left(\textrm{RPL~vs.~CPL}\right) = 0.69 \pm 0.01
\end{equation}
\end{tcolorbox}

\noindent
This value is inconclusive and indicates neither a preference for nor a rejection of the RPL model in comparison to the CPL model. The fact that $\mathcal{B}$ is slightly less than unity can in particular be attributed to the larger dimensionality of the RPL parameter space: compared to the CPL model, the RPL model does not significantly improve the quality of the fit of the NG15 data. At the same time, the RPL model features one parameter more than the CPL model, which results in a larger prior volume and hence a slight suppression of the Bayes factor. In view of these results, we conclude that, at present, the CPL model still suffices to adequately describe the NANOGrav signal. Given the results in Table~\ref{tab:posteriors}, it is, however, conceivable that future PTA data sets may well lead to a measurement of nonzero running. With a bit of luck, $\beta$ may even turn out to be of $\mathcal{O}(1)$, which would be instrumental in the context of model selection.


\newpage
\section{Inflationary gravitational waves}
\label{sec:igw}

Thus far, we only used the RPL model to describe the GWB signal in the PTA frequency band. It is, however, interesting to ask whether the RPL model is also capable of describing a GWB signal extending over a larger frequency range. In the extreme case, one could imagine an RPL-like signal reaching from ultralow frequencies, $f \sim 10^{-(17\cdots16)}\,\textrm{Hz}$, which are probed in CMB observations, to high frequencies, $f \sim 10^{2\cdots3}\,\textrm{Hz}$, which are probed in terrestrial GW interferometer experiments. The physical origin of such a broadband GWB signal would necessarily be primordial. In fact, an RPL-like signal stretching over 20 orders of magnitude in frequency space could originate from cosmic inflation~\citep{Guzzetti:2016mkm}, the stage of accelerated expansion prior to the Hot Big Bang (i.e., the radiation-dominated era in the early Universe). Cosmic inflation stretches primordial scalar and tensor perturbations to super-horizon scales, where they freeze out, before they eventually become dynamical again upon horizon re-entry during radiation domination. Primordial scalar perturbations re-enter the horizon in the form of density perturbations of the primordial plasma, while primordial tensor perturbations turn into propagating GWs upon horizon re-entry. We shall refer to the GWB signal thus produced by inflation as inflationary GWs (IGWs)~\citep{Grishchuk:1974ny,Starobinsky:1979ty,Rubakov:1982df,Fabbri:1983us,Abbott:1984fp}.


\subsection{Upper limits}
\label{subsec:limits}

In this section, we will interpret the RPL model as an approximate phenomenological description of IGWs, which will allow us to supplement the results from our Bayesian fit analysis in the previous section with several observational bounds at lower and higher frequencies. For related earlier work that also interprets the PTA signal in terms of IGWs, see \cite{Vagnozzi:2020gtf,Kuroyanagi:2020sfw,Benetti:2021uea,Vagnozzi:2023lwo}. However, all of these analyses model the IGW spectrum either in terms of a constant power law, a broken power law, or a piecewise power law (in order to account for the possibility of late-time entropy production). The present paper is the first one to consider an RPL-type signal as a simplified model for an IGW spectrum that is also capable of explaining the NANOGrav signal. In total, we are interested in three upper limits on IGWs:

\smallskip\noindent
\textbf{(1) Tensor-to-scalar ratio:} Assuming the RPL-like signal extends from PTA frequencies all the way to CMB frequencies, we must ensure that we do not violate the upper limit on the tensor-to-scalar ratio, i.e., the ratio of the primordial tensor and scalar amplitudes, $r = A_t/A_s$. In order to translate this bound to a constraint on the RPL parameters, we must first map the IGW spectrum onto the RPL model. The GWB spectrum predicted by inflation reads~\citep{Caprini:2018mtu}
\begin{equation}
\label{eq:OmegaIGW}
h^2\Omega_{\rm IGW}\left(f\right) = \frac{h^2\Omega_{\rm rad}^0}{24}\:\mathcal{G}\left(f\right)\mathcal{T}\left(f\right) \mathcal{P}_t\left(f\right) \,.
\end{equation}
Here, $h^2\Omega_{\rm rad}^0 \simeq 4.2 \times 10^{-5}$ is the present-day value of the radiation energy density (assuming three relativistic neutrino species) in units of the critical energy density, times the dimensionless Hubble constant $h$ squared.

The function $\mathcal{G}$ accounts for the varying number of degrees of freedom during the cosmic expansion history,
\begin{equation}
\label{eq:Gfactor}
\mathcal{G}\left(f\right) = \bigg(\frac{g_*\left(f\right)}{g_*^0}\bigg)\bigg(\frac{g_{*,s}^0}{g_{*,s}\left(f\right)}\bigg)^{4/3} \,,
\end{equation}
where $g_*$ and $g_{*,s}$ are the effective numbers of relativistic degrees of freedom entering the radiation energy density $\rho_{\rm rad}$ and radiation entropy density $s_{\rm rad}$, respectively,
\begin{equation}
\rho_{\rm rad} = \frac{\pi^2}{30}\,g_*\,T^4 \,, \qquad s_{\rm rad} = \frac{2\pi^2}{45}\,g_{*,s}\,T^3 \,.
\end{equation}
In order to evaluate these two relations precisely, several effects have to be taken account~\citep{Saikawa:2018rcs,Saikawa:2020swg}, including different quantum statistics for fermions and bosons as well as various perturbative and nonperturbative corrections, which explains why $g_*$ and $g_{*,s}$ typically assume noninteger values. In Eq.~\eqref{eq:Gfactor}, $g_*^0 \simeq 3.38$ and $g_{*,s}^0 \simeq 3.93$ denote the present-day values of these two quantities (assuming three relativistic neutrino species), while $g_*(f)$ and $g_{*,s}(f)$ represent the values of these two quantities in the early Universe when the IGW mode with present-day frequency $f$ and comoving wavenumber $k$ re-entered the Hubble radius, $k = aH$ (with scale factor $a$ and Hubble rate $H$).

The function $\mathcal{T}$ is a transfer function that accounts for the transition from radiation to matter domination,
\begin{equation}
\label{eq:transfer}
\mathcal{T}\left(f\right) = 1 + \frac{9}{16}\bigg(\frac{f_{\rm eq}}{\sqrt{2}\,f}\bigg)^2 \,,
\end{equation}
where $f_{\rm eq} \simeq 2.1 \times 10^{-17}\,\textrm{Hz}$ is the present-day frequency of the IGW mode that re-entered the horizon at the time of matter--radiation equality. Finally, the last factor in Eq.~\eqref{eq:OmegaIGW} is the primordial tensor power spectrum, which describes the strength of GW production during inflation as a function of $f$. We shall assume that inflation gives rise to an RPL-like spectrum of IGWs, such that
\begin{equation}
\mathcal{P}_t\left(f\right) = r\,A_s\left(\frac{f}{f_{\rm CMB}}\right)^{n_t +\sfrac{1}{2}\,\beta_t \ln\left(f/f_{\rm CMB}\right)} \,,
\end{equation}
with tensor-to-scalar ratio $r$, primordial scalar amplitude $A_s \simeq 2.1 \times 10^{-9}$,  primordial tensor index $n_t$, running of the primordial tensor index $\beta_t$, and CMB pivot frequency $f_{\rm CMB} = 0.05\,\textrm{Mpc}^{-1}/(2\pi) \simeq 7.7 \times 10^{-17}\,\textrm{Hz}$.

At the lowest order in the slow-roll parameters, standard single-field slow-roll inflation predicts a consistency relation between the tensor-to-scalar ratio and the primordial tensor index, $n_t = -r/8$~\citep{Liddle:2000cg}, which implies that $n_t$ must be negative. A red-tilted IGW spectrum (i.e., $h^2\Omega_{\rm IGW}$ with $n_t < 0$), however, has no chance of explaining the signal in the PTA band. An important assumption underlying our analysis therefore is that the dynamics of inflation are nonminimal (possibly involving several scalar fields or other particle degrees of freedom), such that the consistency relation $n_t = -r/8$ can be circumvented and a blue-tilted IGW spectrum (i.e., $h^2\Omega_{\rm IGW}$ with $n_t > 0$) is realized. Of course, such a \textit{blue}-tilted signal will still appear as correlated common \textit{red} noise (i.e., $\gamma > 0$) in PTA data, unless $n_t$ is unrealistically large [see Eq.~\eqref{eq:ntRPL} below].

In addition to the factors shown in Eq.~\eqref{eq:OmegaIGW}, we could in principle add another factor, i.e., an additional transfer function $\mathcal{T}_{\rm rh}$ accounting for the transition from reheating to radiation domination. In fact, in~\cite{NANOGrav:2023hvm}, we included precisely such a transfer function. However, in the present paper, we will ignore the dynamics of reheating and simply work with the GWB spectrum in Eq.~\eqref{eq:OmegaIGW} for two reasons: First, the dynamics of reheating are model-dependent and introduce at least three more parameters: the reheating temperature, $T_{\rm rh}$; the equation-of-state parameter during reheating, $w_{\rm rh}$; and the duration of reheating measured in $e$-folds, $N_{\rm rh}$. Second, $\mathcal{T}_{\rm rh}$ only becomes relevant at very high frequencies or for low values of the reheating temperature. Our decision to neglect $\mathcal{T}_{\rm rh}$ thus amounts to the assumption of a high reheating temperature such that $\mathcal{T}_{\rm rh}$ remains irrelevant all the way up to LVK frequencies. 

With Eq.~\eqref{eq:OmegaIGW} at our disposal, we are now ready to match the IGW spectrum to the RPL model in the PTA band. All frequencies in the PTA band are clearly much larger than $f_{\rm eq}$ in Eq.~\eqref{eq:transfer}, which allows us to set $\mathcal{T} = 1$ in our matching procedure. Furthermore, we note that the frequency dependence encoded in $\mathcal{G}$ cannot be captured by the RPL model. This is, however, not a big issue, as $\mathcal{G}$ only varies between $\mathcal{G} = 1$ at low frequencies and $\mathcal{G} \simeq 0.39$ at high frequencies in any case. For the purposes of a rough matching between the IGW and RPL models, it is therefore justified to set $\mathcal{G} = 1$ as well. This leaves us with the simple matching condition
\begin{equation}
\frac{\Omega_{\rm rad}^0}{24}\,\mathcal{P}_t\left(f\right) \approx \frac{2\pi^2}{3H_0^2}\,A^2 f_{\rm ref}^2 \left(\frac{f}{f_{\rm ref}}\right)^{\tilde{n}_{\rm run}\left(f\right)}  \,,
\end{equation}
for frequencies $f$ in the PTA band. Both sides of this condition describe parabola-shaped GWB spectra (if plotted on log--log axes), which allows us to derive a unique solution for the IGW parameters,
\begin{align}
\label{eq:rRPL}
r & = \frac{24}{\Omega_{\rm rad}^0}\frac{1}{A_s} \frac{2\pi^2}{3H_0^2}\,A^2 f_{\rm ref}^2 \left(\frac{f_{\rm CMB}}{f_{\rm ref}}\right)^{\tilde{n}_{\rm run}\left(f_{\rm CMB}\right)} \,, \\
\label{eq:ntRPL}
n_t & = 5 - \gamma - \beta\ln\left(\frac{f_{\rm CMB}}{f_{\rm ref}}\right) \,, \qquad \beta_t = - \beta \,.
\end{align}
We can also think of the matching of the IGW and RPL models as a change of reference frequency, $f_{\rm CMB} \rightarrow f_{\rm ref}$, keeping in mind that $\mathcal{T} \rightarrow 1$ as we move up in frequency space. From this perspective, the relations in Eqs.~\eqref{eq:rRPL} and \eqref{eq:ntRPL} can be understood as a direct consequence of the transformation law in Eq.~\eqref{eq:frefRPL}.

At present, no reliable bounds on $n_t$, let alone $\beta_t$, exist. We therefore only work with the current upper limit on the tensor-to-scalar ratio based on the latest PLANCK and BICEP/Keck data, $r \lesssim 0.03$ at $95\,\%$ C.\,L.\ \citep{Tristram:2021tvh,Galloni:2024lre}, which implies
\begin{tcolorbox}
\vspace{-0.3cm}
\begin{equation}
\label{eq:rbound}
h^2\Omega_{\rm GW}\left(f_{\rm CMB}\right) \lesssim 1.1 \times 10^{-16} \left(\frac{r}{0.03}\right)
\end{equation}
\end{tcolorbox}

\noindent 
This inequality represents the most compact way of writing the constraint on the RPL parameters that follows from the upper limit on $r$. At the same time, it is important to note that, in scenarios where the signal in the PTA band indeed corresponds to an RPL-like spectrum of inflationary origin, the actual, physical amplitude of the GWB spectrum at $f_{\rm CMB}$ is given by $h^2\Omega_{\rm IGW}\left(f_{\rm CMB}\right)$ [see Eq.~\eqref{eq:OmegaIGW}], including the transfer function $\mathcal{T}$, and not by $h^2\Omega_{\rm GW}\left(f_{\rm CMB}\right)$ in the RPL model [see Eq.~\eqref{eq:OmegaRPL}]. The numerical difference between both values is small; but the conceptual difference is worth paying attention to.

\smallskip\noindent
\textbf{(2) Dark radiation:} A primordial GWB from inflation contributes to the energy density of dark radiation (i.e., additional relativistic degrees of freedom beyond the photon and the three Standard Model neutrino species) at the time of BBN. Requiring dark radiation not to spoil the successful BBN prediction of the primordial abundances of the light elements thus puts an upper limit on the integrated energy density of IGWs. This limit can be expressed in terms of $N_{\rm eff}$, the effective number of relativistic neutrino species in the early Universe, or more precisely, $\Delta N_{\rm eff}$, the deviation of $N_{\rm eff}$ from its Standard Model value~\citep{Drewes:2024wbw},
\begin{equation}
\Delta N_{\rm eff} = N_{\rm eff} - N_{\rm eff}^{\rm SM} \,, \qquad N_{\rm eff}^{\rm SM} \simeq 3.0440 \,.
\end{equation}

With this notation, the dark-radiation bound on the IGW energy density reads~\citep{Caprini:2018mtu}
\begin{tcolorbox}
\vspace{-0.3cm}
\begin{equation}
\label{eq:Neffbound}
\int_{\rm f_{\rm BBN}}^{f_{\rm end}}\frac{df}{f}\: h^2\Omega_{\rm GW}\left(f\right) \lesssim 5.6 \times 10^{-6} \Delta N_{\rm eff} 
\end{equation}
\end{tcolorbox}

\noindent Here, $f_{\rm BBN}$ is the present-day frequency of the IGW mode that re-entered the horizon at the onset of BBN at temperatures $T \sim 0.1\,\textrm{MeV}$. For definiteness, we will set $f_{\rm BBN} = 10^{-12}\,\textrm{Hz}$ in what follows. Similarly, $f_{\rm end}$ denotes the present-day frequency of the IGW mode that was just as large as the horizon at the end of inflation. The precise value of $f_{\rm end}$ depends on the dynamics of reheating, in particular, $T_{\rm rh}$, $w_{\rm rh}$, and $N_{\rm rh}$. For large values of the reheating temperature and in the relevant part of the RPL parameter space, the integral over the GW energy density spectrum, however, becomes insensitive to the exact choice of the upper integration boundary. In our analysis, we especially assume that the RPL-like shape of the spectrum persists at least up to the LVK frequency band, which translates to $T_{\rm rh} \gtrsim 10^{10}\,\textrm{GeV}$~\citep{Nakayama:2008wy}. For such large values of $T_{\rm rh}$, we find that the integral over $h^2\Omega_{\rm GW}$ is independent of the exact value of $f_{\rm end}$ to very good approximation. The reason for this is that, for large $f_{\rm end}$ and RPL parameter values satisfying the bound in Eq.~\eqref{eq:Neffbound}, the bulk contribution from the integrand to the total integral is simply located at frequencies $f \ll f_{\rm end}$, such that variations in $f_{\rm end}$ have no numerical impact.

Pictorially speaking, we can say that the RPL spectrum bends away towards smaller values of $h^2\Omega_{\rm GW}$ long before it reaches $f_{\rm end}$. In our numerical analysis, we set $f_{\rm end} = 10^4\,\textrm{Hz}$, for definiteness, and neglect any possible effect of reheating on the shape of the spectrum (see our above comment on the transfer function for the transition from reheating to radiation domination). Moreover, because of the strong suppression of the RPL spectrum as it approaches its endpoint at $f_{\rm end}$ [for parameter values satisfying the bound in Eq.~\eqref{eq:Neffbound}], we conclude that our analysis is less sensitive to possible corrections to the slow-roll dynamics of the inflaton field towards the end of inflation~\citep{Kinney:2021nje}. In fact, assuming a large $T_{\rm rh}$, all relevant frequency scales in our analysis, $f_{\rm CMB}$, $f_{\rm BBN}$, $f_{\rm ref}$, $f_{\rm LVK}$, correspond to IGW modes that exit the horizon long before the end of inflation.

The upper limit on $\Delta N_{\rm eff}$ depends on the choice of cosmological model and combination of data sets. The amount of dark radiation in the early Universe can notably also be constrained by CMB observations, next to the primordial abundances of the light elements. PLANCK data alone yields an upper $95\,\%$ C.\,L.\ limit of around $\Delta N_{\rm eff} \lesssim 0.3$~\citep{Planck:2018vyg}, with the exact value depending on the choice of model fitted to the PLANCK data. However, combining BBN and CMB data, one has to deal with a larger range of uncertainties, which slightly weakens the upper limit~\citep{Pisanti:2020efz,Yeh:2020mgl}. In our analysis, we will hence work with a more conservative bound, $\Delta N_{\rm eff} \lesssim 0.5$.

Before we move on to our third and final parameter constraint, we mention in passing that the integral in Eq.~\eqref{eq:Neffbound} can be solved analytically in the RPL model, 
\begin{equation}
\label{eq:OmegaGW1}
h^2\Omega_{\rm GW}^{\rm tot} = \frac{2\pi^2}{3H_0^2}\,h^2 A^2 f_{\rm ref}^2\,\sqrt{\frac{\pi}{2\beta}}\,e^{\frac{\left(-5+\gamma\right)^2}{2\beta}} \, E_{x_{\rm BBN}}^{x_{\rm end}} \,,
\end{equation}
where we introduced the shorthand notation
\begin{equation}
\label{eq:OmegaGW2}
E_{x_{\rm BBN}}^{x_{\rm end}}  = \left.\textrm{erf}\left(\frac{-5+\gamma+\beta\,x}{\sqrt{2\beta}}\right)\right|_{x_{\rm BBN}}^{x_{\rm end}} \,. 
\end{equation}
Here, $\textrm{erf}$ denotes the Gauss error function and $x_{\rm BBN} = \ln(f_{\rm BBN}/f_{\rm ref})$ and $x_{\rm end} = \ln(f_{\rm end}/f_{\rm ref})$. This analytical expression comes in handy, e.g., when one is interested in varying the integration boundaries $f_{\rm BBN}$ and $f_{\rm end}$ without redoing the whole integral.
Note that, despite the factors of $\sqrt{\beta}$, the formula for $h^2\Omega_{\rm GW}^{\rm tot}$ in Eq.~\eqref{eq:OmegaGW1} is actually also valid for $\beta < 0$. In this case, all imaginary contributions cancel out and the overall result ends up being real and positive, as it should be. Moreover, the expression in Eq.~\eqref{eq:OmegaGW1} also reproduces the correct limit for the case of no running, i.e., the CPL model,
\begin{equation}
\lim_{\beta \rightarrow 0} h^2\Omega_{\rm GW}^{\rm tot} = \frac{2\pi^2}{3H_0^2}\,h^2 A^2 f_{\rm ref}^2\left.\frac{\left(f/f_{\rm ref}\right)^{5-\gamma}}{5-\gamma}\right|_{f_{\rm BBN}}^{f_{\rm end}} \,,
\end{equation}
as well as the correct limit for a flat GWB spectrum
\begin{equation}
\lim_{\substack{\beta \rightarrow 0 \\ \gamma \rightarrow 5}} h^2\Omega_{\rm GW}^{\rm tot} = \frac{2\pi^2}{3H_0^2}\,h^2 A^2 f_{\rm ref}^2\,\ln\left(\frac{f_{\rm end}}{f_{\rm BBN}}\right) \,.
\end{equation}

\smallskip\noindent
\textbf{(3) GWB amplitude at LVK frequencies:} Finally, we require the RPL-like signal from inflation not to violate the upper limit on the amplitude of the stochastic GWB at LVK frequencies~\citep{KAGRA:2021kbb}, 
\begin{equation}
\Omega_{\rm GW}\left(f_{\rm LVK}\right) \lesssim 1.7 \times 10^{-8} \,, \qquad f_{\rm LVK} = 25\,\textrm{Hz} \,.
\end{equation}

\noindent
In the derivation of this limit, the LVK Collaboration set the dimensionless Hubble constant to $h = 0.679$, which implies the following constraint on $h^2\Omega_{\rm GW}$,
\begin{tcolorbox}
\vspace{-0.1cm}
\begin{equation}
\label{eq:LVK}
h^2\Omega_{\rm GW}\left(f_{\rm LVK}\right) \lesssim 7.8 \times 10^{-9} 
\end{equation}
\end{tcolorbox}


\subsection{Discussion}

\begin{figure}
\begin{center}
\includegraphics[width=0.47\textwidth]{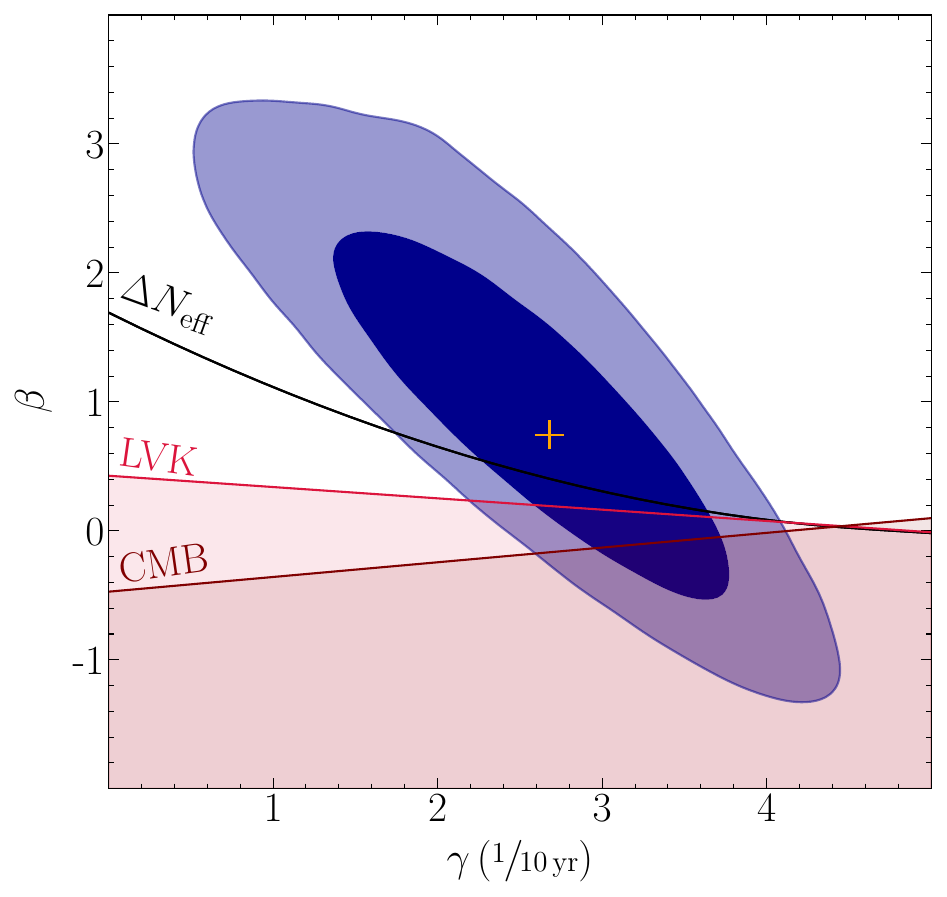}
\end{center}
\caption{Bounds on the RPL parameters $\beta$ and $\gamma$ in scenarios where the NANOGrav signal is identified with an RPL-like signal of inflationary origin. The bounds labeled CMB, $\Delta N_{\rm eff}$, and LVK are derived from the upper limits on the (i) tensor-to-scalar ratio, (ii) amount of dark radiation, and (iii) GWB amplitude at LVK frequencies, respectively. Parameter values below any of the three solid lines are excluded.}
\label{fig:bounds}
\end{figure}

In total, we now derived three constraints on the RPL parameter space [see Eqs.~\eqref{eq:rbound}, \eqref{eq:Neffbound}, and \eqref{eq:LVK}]. In Fig.~\ref{fig:bounds}, we show the three bounds on $\beta$ in dependence of $\gamma$ and for fixed $A$ that result from these constraints. $\log_{10} A$ is fixed at its MAP value, $\log_{10}A = -14.09$ (see Table~\ref{tab:posteriors}), in Fig.~\ref{fig:bounds}, which is justified by the fact that our fit of the RPL model to the NG15 data returns a very narrow credible interval for $\log_{10}A$. That is, $\log_{10} A$ is well constrained by the data, which allows us to reduce our discussion of parameter bounds from a 3D problem to a 2D problem. On top, we observe that the CMB and LVK bounds in Fig.~\ref{fig:bounds} exhibit only a very weak, logarithmic dependence on $A$. Even if we vary $A$ by several orders of magnitude around the value chosen in Fig.~\ref{fig:bounds}, we obtain nearly identical results for the bounds in the $\gamma$\,--\,$\beta$ plane in Fig.~\ref{fig:bounds}, which illustrates the weak sensitivity of our results to uncertainties in the IGW model. The weak, logarithmic dependence of the CMB and LVK bounds on $A$ is a consequence of the large distances on the frequency axis involved in the problem: if we intend to preserve a certain value of $h^2\Omega_{\rm GW}$ at $f = f_{\rm CMB}$ or $f = f_{\rm LVK}$, even a large variation of $A$ at $f = f_{\rm ref}$ can be easily compensated for by a small shift in $\beta$. In short, because of the large hierarchy $f_{\rm CMB} \ll f_{\rm ref} \ll f_{\rm LVK}$, small changes in $\beta$ have a large leverage effect.

This argument does not apply to the $\Delta N_{\rm eff}$ bound, which is not a constraint on a local value of the GWB spectrum, but a global constraint on the integral of the GWB spectrum. Correspondingly, the $\Delta N_{\rm eff}$ is more sensitive to variations of $A$. On the other hand, because of the narrow credible interval for $\log_{10}A$, we are less interested in the behavior of the $\Delta N_{\rm eff}$ bound at $\log_{10}A \gg -14$ or $\log_{10}A \ll -14$. In any case, it would be straightforward to study the $A$ dependence of the $\Delta N_{\rm eff}$ bound using the expressions in Eqs.~\eqref{eq:OmegaGW1} and \eqref{eq:OmegaGW2}.

As evident from Fig.~\ref{fig:bounds}, large regions of the RPL parameter space that can account for the NANOGrav signal are consistent with all three bounds discussed in Sec.~\ref{subsec:limits}. We therefore conclude that a parabola-shaped GWB spectrum of inflationary origin and extending over 20 orders of magnitude in frequency space is indeed a viable option. The fact that only positive values of $\beta$ are in accord with all three bounds in Fig.~\ref{fig:bounds} tells us in particular that this GWB spectrum needs to be negatively curved. That is, while $h^2\Omega_{\rm GW}$ must be an increasing function of frequency in the PTA band (i.e., $\gamma < 5$), the spectrum must eventually bend away towards smaller values again when moving up along the frequency axis, in order to satisfy the $\Delta N_{\rm eff}$ and LVK bounds. Interestingly enough, the CMB bound derived from the upper limit on the tensor-to-scalar ratio actually also allows for positively curved parabolas [i.e., $\beta <0$, which means $\beta_t > 0$; see Eq.~\eqref{eq:ntRPL}]. For instance, an RPL spectrum with $\gamma = 2.5$ and $\beta = -0.1$ could explain the NANOGrav signal and would be consistent with the CMB bound. However, all positively curved parabolas are in conflict with the $\Delta N_{\rm eff}$ and LVK bounds\,---\,as long as we assume that the RPL spectrum does indeed extend all the way up to LVK frequencies and beyond.

For a sufficiently low reheating temperature and hence sufficiently low cutoff frequency in the GWB spectrum, the $\Delta N_{\rm eff}$ and LVK bounds can always be avoided [see our analysis in~\cite{NANOGrav:2023hvm}, in which we also consider the possibility of a very low reheating temperature]. Furthermore, it goes without saying that the bounds in Fig.~\ref{fig:bounds} represent the specific CMB, LVK, and $\Delta N_{\rm eff}$ bounds on the specific model that we are interested in in this paper, the RPL model, if we interpret this model as a simple description of IGWs. From our analysis, we are not able to draw any model-independent conclusions about the dynamics of inflation in general.


Finally, we emphasize that the $\Delta N_{\rm eff}$ and LVK bounds require $\beta$ to be strictly positive. Even the case $\beta = 0$, i.e., the CPL model, is ruled out. This represents an important result of our analysis and a crucial distinction between the minimal CPL model and the next-to-minimal RPL model. If the GWB signal seen in the PTA band should be of inflationary origin, the CPL model would not provide a viable description of this signal that could be extrapolated to very low and high frequencies; the RPL model, on the other hand, does.


\section{Conclusions}
\label{sec:conclusions}

In this paper, we introduced a new model-agnostic template for the spectrum of the GWB signal in the PTA band: the RPL model, which generalizes the simplest spectral template, i.e., the CPL model, by allowing for a logarithmic frequency dependence of the spectral index. In the first part of the paper, we fitted the RPL model to the NG15 data, which provided us with point and interval estimates for the three RPL parameters $A$, $\gamma$, and $\beta$. We find that, at present, the NG15 data is consistent with the assumption of no running of the spectral index, $\beta = 0$, which is reflected in a broad $95\,\%$ credible interval, $\beta \in \left[-0.80,2.96\right]$, and an inconclusive Bayes factor, $\mathcal{B} = 0.69 \pm 0.01$. At the same time, $\beta = 0$ is not contained in its $68\,\%$ credible interval, $\beta \in \left[0.01,1.90\right]$, which nourishes the hope that future PTA data sets may have a chance to find evidence for nonzero $\beta$ and hence deviations from a pure CPL signal. Such a measurement would be instrumental for model selection. In particular, we propose to use the predicted value of $\beta$ in astrophysical and cosmological GWB models as an additional discriminator among different theoretical models. 

In the second part of the paper, we subsequently interpreted the RPL model as a description of an IGW signal, which allowed us to combine the results of our Bayesian fit analysis with upper limits on IGWs at low and high frequencies. Remarkably enough, we found that parabola-shaped GWB spectra of inflationary origin with $\beta > 0$ (i.e., negatively curved spectra) can explain the NANOGrav signal, while at the same time remaining consistent with bounds from BBN, the CMB, and LVK. This is a major success of the RPL model, distinguishing it from the CPL model, for which the same conclusion cannot be drawn. Our results thus motivate theoretical efforts towards the construction of explicit microscopic models of inflation that can achieve the required $A$, $\gamma$, and $\beta$ values identified in this work.

\section*{Acknowledgments}

\smallskip
\centerline{\it Author contributions}
\medskip

This paper uses over a decade's worth of pulsar timing observations and is the product of the work of many people.
K.\,Sc.\ initiated and led the project, designed the analysis, carried out the analytical calculations, and wrote the manuscript.
R.\,R.\,L.\,d.\,S.\ carried out the MCMC analysis and prepared all figures.
D.\,E.\ cross-checked parts of the MCMC analysis.
R.\,v.\,E.\ and T.\,Sc.\ cross-checked and evaluated the parameter bounds in Sec.~\ref{sec:igw}.
The authors thank Andrew Casey-Clyde for suggesting additional references, Timothy Dolch and Ken Olum for helpful feedback and comments on the manuscript, and all members of the NANOGrav New Physics Working Group who provided feedback and comments in working group meetings. 

\bigskip
\centerline{\it Data set}
\medskip

G.\,A., A.\,A., A.\,M.\,A., Z.\,A., P.\,T.\,B., P.\,R.\,B., H.\,T.\,C., K.\,C., M.\,E.\,D., P.\,B.\,D., T.\,D., E.\,C.\,F., W.\,F., E.\,F., G.\,E.\,F., N.\,G.\,D., D.\,C.\,G., P.\,A.\,G., J.\,G., R.\,J.\,J., M.\,L.\,J., D.\,L.\,K., M.\,K., M.\,T.\,L., D.\,R.\,L., J.\,L., R.\,S.\,L., A.\,M., M.\,A.\,M., N.\,M., B.\,W.\,M., C.\,N., D.\,J.\,N., T.\,T.\,N., B.\,B.\,P.\,P., N.\,S.\,P., H.\,A.\,R., S.\,M.\,R., P.\,S.\,R., A.\,S., C.\,S., B.\,J.\,S.\,A., I.\,H.\,S., K.\,St., A.\,S., J.\,K.\,S., and H.\,M.\,W.\ developed timing models and ran observations for the NG15 data set.

\bigskip
\centerline{\it Computing resources}
\medskip

Part of this work was conducted using the High Performance Computing Cluster PALMA II at the University of M\"unster~(\url{https://www.uni-muenster.de/IT/HPC}). The use of the CIS computer cluster at the National Centre for Nuclear Research in Warsaw is gratefully acknowledged.

\bigskip
\centerline{\it Funding information}
\medskip

The NANOGrav Collaboration receives support from National Science Foundation (NSF) Physics Frontiers Center award Nos.\ 1430284 and 2020265, the Gordon and Betty Moore Foundation, NSF AccelNet award No.\ 2114721, an NSERC Discovery Grant, and CIFAR. The Arecibo Observatory is a facility of the NSF operated under cooperative agreement (AST-1744119) by the University of Central Florida (UCF) in alliance with Universidad Ana G.\ M\'endez (UAGM) and Yang Enterprises (YEI), Inc. The Green Bank Observatory is a facility of the NSF operated under cooperative agreement by Associated Universities, Inc. The National Radio Astronomy Observatory is a facility of the NSF operated under cooperative agreement by Associated Universities, Inc. 

\smallskip
L.\,B.\ acknowledges support from the National Science Foundation under award AST-1909933 and from the Research Corporation for Science Advancement under Cottrell Scholar Award No.\ 27553.
P.\,R.\,B.\ is supported by the Science and Technology Facilities Council, grant number ST/W000946/1.
S.\,B.\ gratefully acknowledges the support of a Sloan Fellowship, and the support of NSF under award \#1815664.
M.\,C.\ and S.\,R.\,T.\ acknowledge support from NSF AST-2007993.
M.\,C.\ was supported by the Vanderbilt Initiative in Data Intensive Astrophysics (VIDA) Fellowship.
Support for this work was provided by the NSF through the Grote Reber Fellowship Program administered by Associated Universities, Inc./National Radio Astronomy Observatory.
Pulsar research at UBC is supported by an NSERC Discovery Grant and by CIFAR.
K.\,C.\ is supported by a UBC Four Year Fellowship (6456).
M.\,E.\,D.\ acknowledges support from the Naval Research Laboratory by NASA under contract S-15633Y.
T.\,D.\ and M.\,T.\,L.\ are supported by an NSF Astronomy and Astrophysics Grant (AAG) award number 2009468.
E.\,C.\,F.\ is supported by NASA under award number 80GSFC21M0002.
G.\,E.\,F., S.\,C.\,S., and S.\,J.\,V.\ are supported by NSF award PHY-2011772.
K.\,A.\,G.\ and S.\,R.\,T.\ acknowledge support from an NSF CAREER award \#2146016.
A.\,D.\,J.\ and M.\,V.\ acknowledge support from the Caltech and Jet Propulsion Laboratory President's and Director's Research and Development Fund.
A.\,D.\,J.\ acknowledges support from the Sloan Foundation.
The work of N.\,La., X.\,S., and D.\,W.\ is partly supported by the George and Hannah Bolinger Memorial Fund in the College of Science at Oregon State University.
N.\,La.\ acknowledges the support from Larry W.\ Martin and Joyce B.\ O'Neill Endowed Fellowship in the College of Science at Oregon State University.
Part of this research was carried out at the Jet Propulsion Laboratory, California Institute of Technology, under a contract with the National Aeronautics and Space Administration (80NM0018D0004).
R.\,R.\,L.\,d.\,S.\ is supported in part by the National Science Centre (Poland) under the research Grant No.\ 2020/38/E/ST2/00126.
D.\,R.\,L.\ and M.\,A.\,M.\ are supported by NSF \#1458952.
M.\,A.\,M.\ is supported by NSF \#2009425.
C.\,M.\,F.\,M.\ was supported in part by the National Science Foundation under Grants No.\ NSF PHY-1748958 and AST-2106552.
A.\,Mi.\ is supported by the Deutsche Forschungsgemeinschaft under Germany's Excellence Strategy\,--\,EXC 2121 Quantum Universe\,--\,390833306.
The Dunlap Institute is funded by an endowment established by the David Dunlap family and the University of Toronto.
K.\,D.\,O.\ was supported in part by NSF Grant No.\ 2207267.
T.\,T.\,P.\ acknowledges support from the Extragalactic Astrophysics Research Group at E\"{o}tv\"{o}s Lor\'{a}nd University, funded by the E\"{o}tv\"{o}s Lor\'{a}nd Research Network (ELKH), which was used during the development of this research.
H.\,A.\,R.\ is supported by NSF Partnerships for Research and Education in Physics (PREP) award No.\ 2216793.
S.\,M.\,R.\ and I.\,H.\,S.\ are CIFAR Fellows.
Portions of this work performed at NRL were supported by ONR 6.1 basic research funding.
J.\,D.\,R.\ also acknowledges support from start-up funds from Texas Tech University.
The work of K.\,Sc., T.\,S., and R.\,v.\,E.\ is supported by Deutsche Forschungsgemeinschaft (DFG) through the Research Training Group (Graduiertenkolleg) 2149: Strong and Weak Interactions\,--\,from Hadrons to Dark Matter. 
J.\,S.\ is supported by an NSF Astronomy and Astrophysics Postdoctoral Fellowship under award AST-2202388, and acknowledges previous support by the NSF under award 1847938.
C.\,U.\ acknowledges support from BGU (Kreitman fellowship), and the Council for Higher Education and Israel Academy of Sciences and Humanities (Excellence fellowship).
C.\,A.\,W.\ acknowledges support from CIERA, the Adler Planetarium, and the Brinson Foundation through a CIERA-Adler postdoctoral fellowship.
O.\,Y.\ is supported by the National Science Foundation Graduate Research Fellowship under Grant No.\ DGE-2139292.


\bibliography{arXiv_v2.bib}{}
\bibliographystyle{aasjournal}
\end{document}